\documentclass[submission, Phys]{SciPost}

\usepackage{times,bbm}
\usepackage{xcolor}
\usepackage{mathtools}

\usepackage{graphicx}
\usepackage{float}
\graphicspath{{./Figures/}}

\usepackage{hyperref}
\hypersetup{
	breaklinks,
	colorlinks,
	linkcolor=black,
	citecolor=black,
	urlcolor=black,
}

\definecolor{jens}{rgb}{.2,0.7,.9}

\newtheorem{measure}{Measure}

\newcommand{\Tr}{\text{Tr}}
\newcommand{\ket}[1]{\left \vert #1 \right \rangle}
\newcommand{\bra}[1]{\left \langle #1 \right \vert}

\newcommand{\intd}{\mathrm{d}}

\begin{document}

\begin{center}{\Large \textbf{
Experimentally accessible witnesses of many-body localization
}}\end{center}

\begin{center}
M.\,Goihl$^{1,*}$, M.\,Friesdorf$^1$, A.\,H.\,Werner$^{1,2}$, W.\,Brown$^{1,3}$ and
J.\,Eisert$^1$
\end{center}

\begin{center}
$^1$ Dahlem Center for Complex Quantum Systems, Freie Universit{\"a}t Berlin, 14195 Berlin, Germany

$^2$ Department of Mathematical Sciences, University of Copenhagen, DK-2100 K{\o}benhavn, Denmark

$^3$ Northrop Grumman Corporation, Baltimore, MD 21240, USA

* mgoihl@physik.fu-berlin.de
\end{center}


\section*{Abstract}
{\bf
The phenomenon of many-body localized (MBL) systems has attracted significant interest in recent years, for its intriguing implications from a perspective of both condensed-matter and statistical physics: they are insulators even at non-zero temperature and fail to thermalize, violating expectations from quantum statistical mechanics. What is more, recent seminal experimental developments with ultra-cold atoms in optical lattices constituting analog quantum simulators have pushed many-body localized systems into the realm of physical systems that can be measured with high accuracy. In this work, we introduce experimentally accessible witnesses that directly probe distinct features of MBL, distinguishing it from its Anderson counterpart. We insist on building our toolbox from techniques available in the laboratory, including on-site addressing, super-lattices,
and time-of-flight measurements, identifying witnesses based on fluctuations, density-density correlators, densities, and entanglement. We build upon the theory of out of equilibrium quantum systems, in conjunction with tensor network and exact simulations, showing the effectiveness of the
tools for realistic models.}

\vspace{10pt}
\noindent\rule{\textwidth}{1pt}
\tableofcontents\thispagestyle{fancy}
\noindent\rule{\textwidth}{1pt}
\vspace{10pt}

\section{Introduction}
Many-body localization provides a puzzling and exciting paradigm within quantum many-body physics and is for good reasons attracting
significant attention in recent years. Influential theoretical work \cite{Basko} building upon the seminal insights by Anderson
on disordered models \cite{Anderson} suggested that localization would survive the presence of interactions. Such many-body localized models, as
they were dubbed, would be insulators even at non-zero temperature and exhibit no particle transport. Maybe more strikingly from the
perspective of statistical physics, these many-body localized models would fail to thermalize
following out of equilibrium dynamics \cite{HuseReview,PalHuse,Oganesyan},
challenging common expectations how systems ``form their own heat bath'' and hence tend to be locally well described by the familiar canonical Gibbs ensemble \cite{Polkovnikov_etal11,1408.5148,christian_review}. Following these
fundamental observations, a ``gold rush'' of theoretical work followed,
identifying a plethora of phenomenology of such many-body localized models. They would exhibit a distinct and peculiar
logarithmic scaling of entanglement in time \cite{Prosen_localisation,Pollmann_unbounded},
the total correlations of time averages have a distinct scaling \cite{1504.06872},
many Hamiltonian eigenstates fulfil area laws \cite{AreaReview} for the entanglement entropy
\cite{Bauer,1409.1252} and hence violate what is called the eigenstate thermalization hypothesis \cite{Srednicki94}. The
precise connection and interrelation between these various
aspects of many-body localization is just beginning to be understood \cite{1409.1252,1412.3073,1410.0687,1412.5605,SerbynPapicAbanin,huse2014phenomenology}, giving
rise to a vivid discussion in theoretical physics.

These theoretical studies have recently been complemented by seminal experimental activity, allowing to probe models that are
expected to be many-body localized in the laboratory under remarkably controlled conditions \cite{BlochMBL,1509.00478}. 
This work goes much beyond earlier demonstrations
of Anderson localization in a number of models \cite{AndersonLight}, in that now actual interactions are expected to be relevant. Such ultra-cold atomic systems indeed
provide a pivotal arena to probe the physics that is at stake here \cite{BlochSimulation}. What is still missing, however, is a direct detection of
the rich phenomenology of many-body localization in the laboratory. Rather than seeing localization and taking the presence of interactions for granted,
it seems highly desirable to make use of these novel exciting possibilities to directly see the above features, distinctly separating the observations
from those expected from non-interacting Anderson insulators. Such a mindset is that of ``witnessing'' a property, somewhat inspired by how
properties like entanglement are witnessed \cite{quant-ph/0607167,Audenaert06,Guehne} in quantum information.

In this work, we aim at capturing precisely those aspects of the rich phenomenology
of many-body localization that are directly accessible with present experimental tools. We
would like to provide a ``dictionary'' of possible tools,  as a list or a classification of features that can be probed
making use of only in situ site resolved measurements, including the
measurement of density-density correlations and
time of flight measurements, in conjunction with a variation of densities. In this way, we aim at identifying a comprehensive
list of features that ``could be held responsible'' for MBL, based on data alone.
While all we explicitly state is directly related to cold atoms in optical lattices, a similar approach is expected to be feasible
in continuous cold bosonic atoms on atom chips \cite{SchmiedmayerScience,cMPSTomographyShort},
where correlation functions of all orders can readily be directly
measured. We leave this  as an exciting perspective.

\section{Probing disordered optical lattice systems}

The setting we focus on is that of interacting (spin-less) fermions
placed into a one-dimensional optical lattice, a setting that prominently allows
to probe the physics under consideration \cite{BlochMBL,BlochSimulation}.
Such systems are well described by
\begin{equation}
  \label{eq:dis_ham}
  H = \sum_j \left( f_j^\dagger f_{j+1} + \text{h.c.} \right) + \sum_j w_j n_j + U \sum_j n_j n_{j+1}  ,
\end{equation}
where $f_j$ denotes a fermionic annihilation operator on site $j$ and
$n_j=f_j^\dagger f_j$ is the local particle number operator.
The disorder in the model is carried by the local potential-strength $w_j$, which is
drawn independently at each lattice site $j$ according to a suitable
probability distribution. Experimentally, the disorder can either be realized by superposing
the lattice with an incommensurate laser or by speckle patterns \cite{BlochMBL}.
From Eq.\ \eqref{eq:dis_ham} one obtains the disordered
Heisenberg chain \cite{Laflorencie2015}
by setting $U=2$ and scaling the disorder by a factor two. To keep the discussion
conceptually clear, just as Ref.\ \cite{Laflorencie2015} we make use of a uniform
distribution on the interval $[-I,I]$, where we refer to $I$ as
the disorder strength. Thus, for $U=2$ the ergodic to MBL phase transition is  approximately at $I \approx 7$ \cite{Laflorencie2015}. 
Most of the known experiments of MBL have been carried out in a related model
of on-site interacting bosons for which we show data in Appendix
\ref{app:bosons}.

The phase diagramme of these models is best known for
$U=0$ corresponding to the non-interacting Anderson insulator
and for $U=2$, the MBL phase. To add a flavour of usual phase transitions
order parameters such as total correlations \cite{1504.06872},  
fluctuations of local observables \cite{Pollmann2015} or
the structure of the eigenstates \cite{PhysRevX.5.041047} have been suggested.
While these quantities impressively signal the transition, it is not a priori
clear whether they can be implemented in an actual experiment.
Recent numerical studies\cite{SerbynMBL} show that pump-probe type setups
and novel instances of spin noise spectroscopy \cite{MoessnerMBL} as well as
utilizing MBL systems as a bath \cite{Vasseur} are
indeed suited to distinguish the above phases, albeit experimental realizations of this
endeavour appear to need substantial changes and innovations in realistic setups.
Another possibility for the phase distinction, which has prominently been carried out experimentally \cite{1509.00478},
is given by observing the behaviour of quasi two-dimensional systems in
comparison to their one dimensional counterparts.
While this impressively demonstrates the capabilities of optical lattices as platforms for quantum simulations,
it does not test the properties of MBL in one dimension as such.
We set out to find comparably strong and direct signatures of one dimensional MBL, 
which however rely on simple established measurement operations. Hence, we start by summarizing the measurements, which
we conceive to be feasible in an optical lattice experiment.

\section{Measurements considered feasible}
We now turn to specifying what measurements we consider feasible in
optical lattices with state of the art techniques. For this,
we focus on the following two types of measurements:

\textbf{In-situ}: An in-situ measurement detects the
occupation of
individual lattice sites. This technique only allows to resolve the parity
of the particle
number on each site, which for fermions constitutes no limitation, however.
Using the fact that single-shot measurements are performed, also higher
moments like density-density correlators
can be extracted from this kind of measurements. Both ramifications will be
used. This measurement has been used to determine onsite parities
in Ref.\,\cite{Choi1547} to show particle localization in two dimensional disordered optical
lattices. Here, we try to additionally witness the
interactions necessary to distinguish Anderson from MBL systems.

\textbf{Time-of-flight}: The time-of-flight (ToF)
measurement extracts position-averaged
momentum information of the form
\begin{equation*}
  \langle n(q,t_{\text{ToF}}) \rangle = |\hat{w}_0(q)|^2 \sum_{j,k}
  e^{i q (r_j- r_k) - i
  \frac{c (r_j^2 + r_k^2)}{t_{\text{ToF}}}} \langle f_j^\dagger f_k \rangle \;,
\end{equation*}
where $\{r_j\}$ are the positions of the lattice sites,
$\hat{w}_0$ reflects the Wannier functions in momentum space,
and $c>0$ is a constant derived from the mass of the particles and the lattice
constant. This measurement was used in Ref.\,\cite{BlochMBL} to 
determine the imbalance -- a measure of localization. 

The main goal of this work is to identify key quantities that indicate that the system
indeed is many-body localized based on measurement
information extracted using these two techniques.
Here, we want to show both that the system is
localized and that it is interacting. Thus, we also
want to convincingly detect the difference between an MBL system
and a non-interacting Anderson insulator.
In order to approach this task, we will look at the time
evolution of an initial state that is particularly easy
to prepare experimentally relying on optical super-lattices \cite{BlochMBL,Trotzky_etal12},
namely an alternating pattern of the form 
\begin{align}
  \label{initial_state}
  \ket{\psi(t=0)} = |0,1,0,1, \cdots0,1 \rangle \;.
\end{align}
This initial product state will, during time evolution,
build up entanglement and become correlated \cite{Prosen_localisation,Pollmann_unbounded}. 
Naturally, this is far from being the only choice for an initial state and
alterations in this pattern and correspondingly locally changing particle and hole
densities would surely be insightful, specifically since a modulation of the density 
already points towards interactions in the MBL phase being significant.
In this work, we put emphasis 
on measurements although preparation procedures as the above mentioned density
variations are an interesting problem in its own right. However, as we will demonstrate,
the above defined initial state already captures the colourful phenomenology of MBL in all of its salient aspects.

\begin{figure}[t]
\centering
  \includegraphics[width=0.6\textwidth]{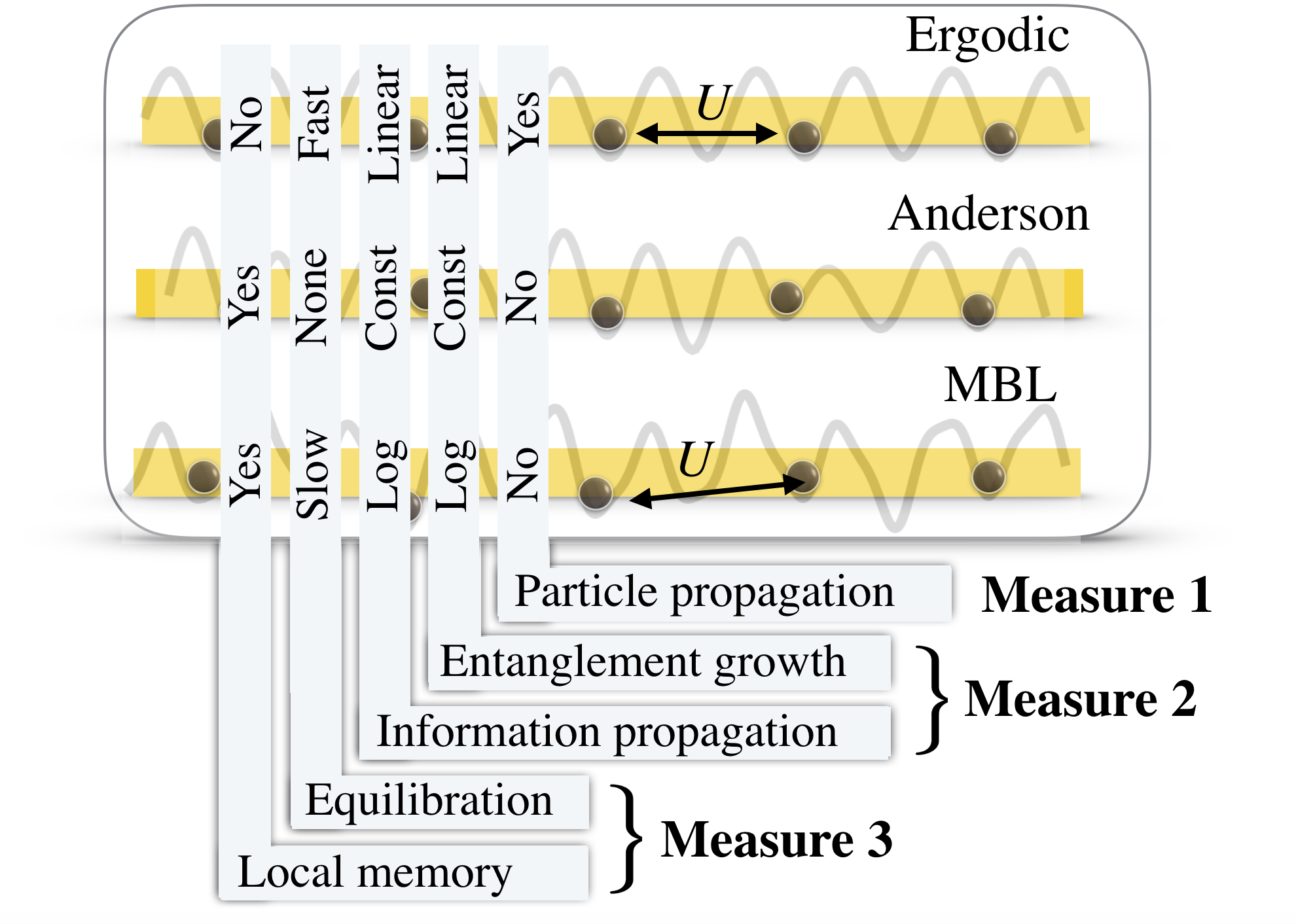}
  \caption{An overview over the dynamical behaviour of MBL systems 
  versus their ergodic and thermalizing and Anderson localized
  counterparts. \textbf{Measure 1} detects particle 
  propagation and phase correlations
  and can be implemented using ToF imaging. \textbf{Measure 2} and \textbf{Measure 3}
  utilise in-situ imaging to observe density-density correlations and equilibration
  behaviour.}
\end{figure}

\begin{figure*}[t]
  \centering
  \includegraphics[width=\textwidth]{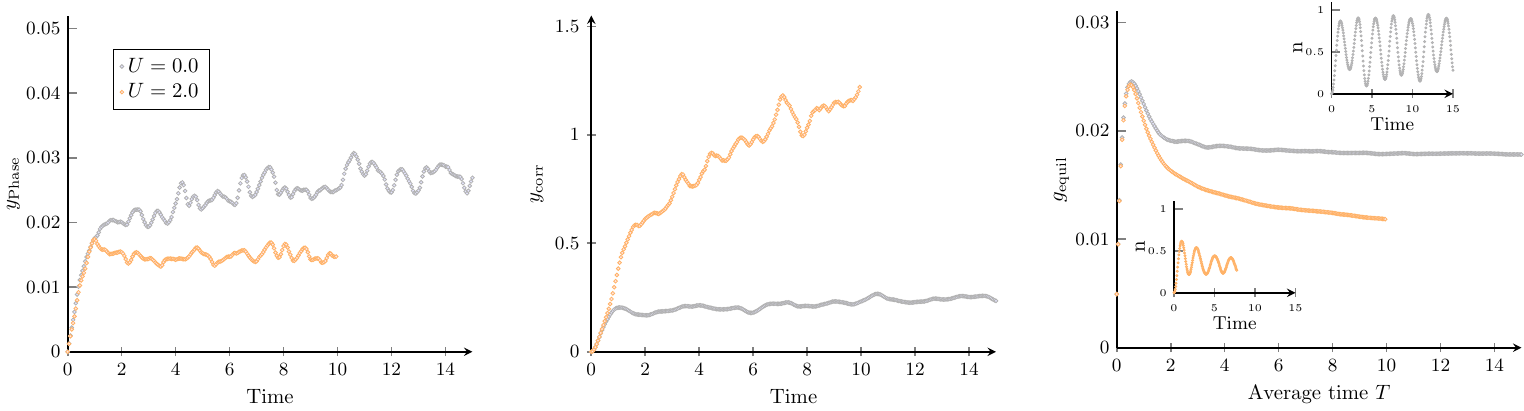}
  \caption[Numerical findings]{
    Plotted are the results of a TEBD simulation \cite{opentebd} 
    of the dynamical evolution of the initial state $\psi$ from Eq.~\eqref{initial_state}
    under the Hamiltonian in Eq.~\eqref{eq:dis_ham} for the case of an Anderson insulators with $U=0$ and
    MBL with $U=2$. The disorder strength is $I=8$.
    The three plots are averaged over 100 disorder realizations.
    \textbf{Left:} 
    Shown is the time evolution of $y_{\text{Phase}}$ defined in \textbf{Measure 1} 
    demonstrating that the phase correlation behaviour saturates both for 
    MBL and Anderson localization.
    \textbf{Center:}
    The plot shows the dynamical evolution of $y_{\text{Corr}}$ defined in \textbf{Measure 2}.
    Information propagation is fully suppressed in an Anderson insulator, resulting in a saturation
    of this quantity. In contrast, correlations continue to spread in the MBL system
    beyond all bounds, giving rise to a remarkably strong signal feasible to be detected in
    experiments.
    \textbf{Right:}
    Shown are the averaged fluctuations $g_{\mathrm{Eq}}$ defined in \textbf{Measure 3}
    as a function of the time $T$ over which the average is performed.
    The insets show the time evolution of the particle density at the position $L/2$, which enters the 
    calculation of $g_{\mathrm{Eq}}$ for one disorder realization which is identical for the 
    MBL and Anderson localized model.
    As the insets also show, the local fluctuations continue indefinitely for the Anderson insulator,
    corresponding to a saturation of $g_{\mathrm{Eq}}$, while the MBL system equilibrates and
    $g_{\mathrm{Eq}}$ continues to decrease accordingly.
  }
  \label{mainfig}
\end{figure*}

\section{Phenomenology of many-body localization}
A fundamental characteristic of MBL is the presence of local constants of motion \cite{HuseReview}.
They are approximately local operators $\widetilde\sigma^z_j$ whose support is centred on lattice site $j$,
but which nevertheless commute with the Hamiltonian, i.e., $[H,\widetilde\sigma_j]=0$.
These operators are mainly supported on a region with diameter $\xi$, corresponding to the localization length scale of the system.
In fact, in the MBL regime, the dynamics can be captured by a phenomenological model in terms of a set of mutually commuting
quasi-local constants of motion,
\begin{align}
\label{iomh}
  H^{(2)}_{l-\mathrm{Bit}} = \sum_i \mu_i \tilde \sigma^z_i + \sum_{j<i} J_{i,j} \tilde \sigma^z_i \tilde \sigma^z_j \; .
\end{align}
Such a Hamiltonian constitutes a second order approximation of what is 
 known as the $l$-bit model \cite{SerbynPapicAbanin,huse2014phenomenology,Ros} to exemplify the dynamics.
Here $\tilde{\sigma^z_j}$ again denotes a quasi-local integral of motion centred
on site $i$, $\mu_i$ is a random onsite potential and the coupling strength $J_{i,j}$ between constants of motion is assumed to decay
 suitably fast in their distance
$d(i,j)$.
In particular, it is expected that the dynamics generated by the Hamiltonian defined in Eq.~\eqref{eq:dis_ham}
 in the MBL regime corresponding to $U=2$ and $ I>7$ can be well captured by the $l$-bit model.

This phenomenological model gives rise to a separation of time scales in the
evolution into two regimes.
Initially, there is a fast regime, where the evolution
takes place mainly inside the support of each
local constant of motion $\tilde \sigma^z_i$.
Hence, for this time scale, transport is unconstrained and
particles and energies can move freely inside the localization length.
Correspondingly, information can spread ballistically.
Beyond the localization length, the dynamics is dominated by the coupling of the
constants of motion, given by the second term in
Eq.~\eqref{iomh} \cite{huse2014phenomenology}.
The intuition is that this evolution does not facilitate particle or
energy propagation, leading to a complete break-down of
thermal and electric conductivity.
Nevertheless, the couplings between distant constants of motion
allow for the creation of correlations over arbitrary length scales given sufficient time.
This dephasing mechanism in turn makes it possible to send information and
yields an explanatory mechanism for the observed slow growth of entanglement \cite{Prosen_localisation,Pollmann_unbounded,1412.3073},
measured as the von Neumann entropy of the half chain of an
infinite system $S(t) = \Theta(\log(t))$ (in Landau notation).

Mathematically, these two dynamical regimes are best distinguished by the
effect of a local unitary
excitation on distant measurements.
More precisely, given a local measurement $O_A$ supported in a spatial region $A$ and a
unitary $V_B$ corresponding to a local excitation in a region $B$,
we wish to bound the change in expectation value of $O_A(t)$ induced
by the unitary excitation.
This can be cast into a Lieb-Robinson bound \cite{LiebRobinson72,1409.1252} of the form
\begin{eqnarray}
\label{LR}
  \left|\langle V_B O_A(t) V_B^\dagger \rangle - \langle O_A(t) \rangle\right|
  \leq C(A)
  \begin{cases}
    e^{-\mu (d(A,B) - v |t|)} \quad \text{\textbf{ I}}, \\
    t e^{-\mu d(A,B)} \quad \text{\textbf{ II}},
  \end{cases}
\end{eqnarray}
where $C(A)$ a constant depending on the support of $O_A$.
For the connection between different zero velocity Lieb-Robinson bounds
and the necessity of a linear $t$-dependence in \textbf{II},
see Ref.\ \cite{1409.1252}.
Here $\textbf{I}$ corresponds to the ballistic regime and $\textbf{II}$ captures 
the slower dephasing. 
In the context of optical lattices, local excitations seem difficult
to implement. 
Hence, in the following, we focus on the observation
of indirect effects on the dynamical evolution in MBL systems.

\section{Feasible witnesses}

In the following, we demonstrate that local memory of initial conditions, slow spreading of correlations and equilibration
of local densities provide clear measures to distinguish MBL systems from both the non-interacting Anderson insulators
and also from ergodic systems, i.e., those where local measurements,
after a short relaxation time, can be captured by thermal ensembles.
In order to carry out our analysis, we will complement the intuitive guideline provided by the phenomenological $l$-bit model
by a numerical tensor network TEBD simulation \cite{Daley2004} (for details see Appendix \ref{app:numerics}). 
The chosen parameters for the simulation are a disorder strength of $I=8$ and interaction strengths of $U=2$ or $U=0$
for the MBL and Anderson case, respectively.
We begin by considering the influence of the suppression of particle propagation.

\subsection{Absence of particle transport}
  A defining feature of localized systems is that independent of the interaction
  strength, particles and energies do not spread over the entire system, but remain
  confined to local regions.
  They merely redistribute inside the localization length, which can be
  extracted from the constants of motion. Therefore, even for long
  times, the particle density profile of an MBL system will not move to
  its thermal form,
  but rather retain some memory of its initial configuration.
  This gives rise to the following particle localization measure.
  \begin{measure}[Particle propagation and phase correlations]
  We define the following measure 

  $y_{\mathrm{Phase}}(t)$, which probes particle propagation
  for a system of length $L$
  \begin{align}
    f_{\mathrm{Phase}}(k,t) &:= \left | \langle f_{L/2}^\dagger(t) f_{L/2+k}(t) \rangle \right|, \\
    y_{\mathrm{Phase}}(t) &= \sum_k f_{\text{Phase}}(k,t) k^2 \; .
  \end{align}
  \end{measure}
  On an intuitive level, this measure directly probes the spread of particles,
  including weights based on the distance to the initial position $L/2$ such that
  distant contributions are amplified.
  
  Numerically, we find that $y_{\text{phase}}(t)$ initially shows a steep linear increase, 
  indicative of the ergodic dynamics governed by the onsite term of Eq.\ \eqref{iomh} (Fig.~\ref{mainfig}). 
  In the second regime it fluctuates without visible growth, indicating a break-down of
  particle transport on length scales beyond the localization length.
  Thus, the length scale of the phase correlations established
  in the system can be bounded independent of time $y_{\text{Phase}} (t) = O(1)$.
  For ergodic systems, where particles and energies spread ballistically,
  the measure would grow in an unconstrained fashion over time.
  Based on this insight, we deduce that time-of-flight images,
  while clearly distinguishing between localized and ergodic phase,
  are not useful for the distinction between interacting and non-interacting localized
  systems.

  Again more formally, this measure can be understood by considering the time evolution
  of the correlation matrix given by the matrix elements
  \begin{align}
    \gamma_{j,k} (t) &\coloneqq \langle f_j^\dagger(t) f_k(t) \rangle ,
  \end{align}
  where $ \langle f_j^\dagger f_k \rangle = \Tr ( f_j^\dagger(t) f_k(t) \rho) )$.
  For the non-interacting case of an Anderson insulator, this evolution is unitary $\gamma (t) = U(t) \gamma (0) U^\dagger(t)$,
  where $ f_j^\dagger (t) = \sum_l U_{j,l}(t) f_l^\dagger$ is the evolution of the fermionic mode operators.
  For an Anderson insulator, dynamical localization precisely corresponds to locality of the unitary evolution \cite{0709.3707},
  meaning that the matrix elements of $U$ are expected to decay exponentially $|U_{j,k}(t)| \leq C e^{-d(j,k)}$ for some constant $C$
  with high probability \cite{AndersonDecayBounds}.

  In the case of interacting Hamiltonians that conserve the particle number, this time evolution can be captured
  in form of a quantum channel  
  \begin{align}
    \gamma (t) = \sum_{l=1}^{L^2} K_l(\rho_0, t) \gamma(0) K_l^\dagger(\rho_0, t) ,
  \end{align}
  where the Kraus operators $K_l(\rho_0, t)$ depend on the full initial state.
  As particle propagation in an MBL system is expected to also be localized, it is assumed that the individual Kraus operators obey
  $|K_{j,k}(\rho_0, t)| \leq C_K e^{-d(j,k)}$.
  Starting from an initial product state of the form in Eq.~\eqref{initial_state}, we obtain
  \begin{align}
    \gamma_{z_1,z_2}(t) &= \langle f_{z_1}^\dagger(t) f_{z_2}(t) \rangle\nonumber \\
    &= \sum_{j,l} \bra{z_1} K(\rho_0,t) \ket{j} \gamma(0)_{j,l}
    \bra{l} K(\rho_0,t) \ket{z_2} \nonumber\\
    &= \sum_{j \, \mathrm{even}} C_K^2 e^{- d(j,0)- d(z_1+j,z_2)}.
  \end{align}
  This again results in a suppression with the distance between $z_1$ and $z_2$, 
  causing a saturation of the phase correlation measure 
  $f_{\text{Phase}}(k,t)$ independent of time.

\subsection{Slow spreading of information}

  While particles and energies remain confined in interacting
  localized systems, correlations
  are expected to show an unbounded increase over time
  \cite{Pollmann_unbounded,Prosen_localisation}, although slower than in the ergodic
  counterpart. In stark contrast, Anderson localized many-body systems will
  not build up any correlations that go beyond the localization length.
  In order to probe the spreading of correlations in the system, we focus
  on a quantity easily accessible in the context of optical lattices,
  using in-situ images for different evolution times.
  As it turns out, this kind of simple density-density correlator is
  already sufficient to separate Anderson localization from MBL systems.

  \begin{measure}[Logarithmic information propagation] In order to examine the
  spatial spreading of density-density correlations, we define
  the quantity $y_{\mathrm{Corr}}(t)$,
  \begin{eqnarray}
    f_{\mathrm{Corr}}(k,t) &:=&   | \langle n_{L/2} n_{L/2+k} \rangle - \langle n_{L/2} \rangle \langle n_{L/2+k} \rangle |,\\
    y_{\mathrm{Corr}}(t) &:= & \sum_k f_{\mathrm{Corr}}(k,t) k^2.
    \end{eqnarray}
  \end{measure}
  $y_{\mathrm{Corr}}$ is a direct indicator
  for the length scale over which density-density correlations are
  established without having to resort
  to assuming an explicit form, such as a decay in terms of an exponential function.

  Comparable to the dynamics of the phase correlations, we numerically find a steep
  initial increase followed by a saturation for the non-interacting case (Fig.~\ref{mainfig}).
  The MBL system however, continues to build up density-density correlations for
  the times simulated. There is a transition in propagation speed, which we
  ascribe to the two dynamical regimes discussed before. Hence we conclude,
  that density-density correlations can be used to discriminate MBL from its
  non-interacting counterpart. 

  An intuitive explanation for the spread of density-density correlations despite
  spatial localization of particles is that after exploring the localization length,
  the particles feel the presence of neighbouring particles. 
  Mediated by this interaction, the local movement of particles, governed by the respective constant of motion,
  becomes correlated, even over large distances.
  In contrast, in the Anderson insulator where constants of motion are completely decoupled, this communication cannot
  take place.

  We can connect this intuitive explanation to the more rigorous setting
  of Lieb-Robinson-bounds.
  In the Anderson insulator in one-dimension, there provably exists a
  zero-velocity Lieb-Robinson-bound, where the correlator on the left
  hand side of Eq.~\eqref{LR} is bounded by a time independent factor $e^{-\mu d(A,B)}$.
  This means that the detectability of an excitation created in region $A$
  decreases exponentially with the distance to $B$.
  On the contrary, in the MBL regime we expect a
  logarithmic Lieb-Robinson cone of the form
  of Eq.~\eqref{LR} \textbf{II}.
  Hence, an unbounded growth of correlations between distant regions is
  in principle possible, given sufficient time. Furthermore, we have shown
  that this built-up
  of correlations does also happen on observable time scales as can
  be seen from the evolution of density-density correlations captured
  by \textbf{Measure 2}.

\subsection{Dephasing and equilibration}
  It is also instructive to study the differences between the Anderson and MBL-regime with
  respect to their equilibration properties. Due to the interactions present,
  we expect equilibration of fluctuations to take place in MBL systems, whereas in Anderson
  insulators the effective subspaces explored by single particles remain
  small for all times and hence fluctuations remain large. 
  This in turn implies that fluctuations of local expectation
  values die out in the interacting model, but persist in an Anderson insulator.
  This qualitative difference has already been identified as a signifier of interactions
  in a disordered system\cite{BlochMBL}. Here, we build upon this idea and
  propose to consider the average change rate of local expectation values
  in order to detect the decreasing fluctuations in the MBL phase.

  \begin{measure}[Density evolution: Equilibration of fluctuations]
    \label{def:derivative}
    We consider the expectation value $f_{\mathrm{Eq}}(t) = \langle n_{L/2}\rangle(t)$ of a local
    density operator in the middle of the system.
    As a measure of local equilibration, we introduce the
    averaged rate of change of this density as a function of time $T>0$
    \begin{align}
      g_{\mathrm{Eq}}(T) := \frac{1}{T} \int \limits_0^T \intd t \, |f_{\mathrm{Eq}} '(t)| .
    \end{align}
  \end{measure}
  As laid out in Fig.\ \ref{mainfig} again,  this function over time indeed shows a remarkably smooth behaviour
  that allows for the clear distinction between
  an Anderson localized system and its MBL counterpart in that after a mutual increase
  the Anderson system saturates at a constant value, whereas in the MBL phase,
  $g_{\mathrm{Eq}}$ shrinks successively.

  If we again resort to the Lieb-Robinson bound picture, we find that in the
  Anderson case a local excitation is confined to a distinct spatial region
  given by the zero-velocity Lieb-Robinson-bound introduced in the previous
  chapter. This implies that the effective subspace explored is constant and
  specifically, the excitation cannot build up long distance correlations
  and fluctuations remain large.
  This can also be seen from the results of \textbf{Measure 2}.
  If we now, however, turn to the interacting model, a local excitation
  will slowly explore larger and larger parts of the Hilbert space, leading
  to a slow, but persistent decrease of the fluctuations.

Recently, there was an impressive progress in measuring quantities
very much related to the entanglement entropy in small
one-dimensional optical lattices \cite{greiner1,greiner2}. In both
of these works, quantities alike to our \textbf{Measure 2} are used
as well. In Ref.\,\cite{greiner2}, the authors define a quantity called
transport distance which basically coincides with our \textbf{Measure 2}.
The difference being that their scaling function is only linear instead of
quadratic. However, they do not employ this measure to show the many-body
correlations in these systems. Rather, they calculate number and 
configurational entanglement \cite{greiner1}. The system sizes used are
very restricted, possibly due to the complicated procedure of obtaining these
entropies. We think that an implementation of \textbf{Measure 2} or
\textbf{Measure 3} might complement these results nicely 
by overcoming these problems and hence being applicable also for larger
systems and potentially also higher dimensional systems, where
the fate of MBL is still debated.

\section{Conclusions and outlook}

In this work, we proposed an operational procedure for distinguishing
MBL phases building upon realistic measurements, which can be performed
in the realm of optical lattices with present technology. Utilising a phenomenological model
and the concept of Lieb-Robinson-bounds, we explained the effects numerically
investigated employing tensor network methods.
The equilibration of local observables allows for the distinction
of Anderson and MBL localized models. Density-density correlations
allow for the same information bit extraction, while furthermore reproduce
the expected phenomenology. Further investigating this quantity might
yield information about the localization length via the duration of
the first evolution regime.

Phase correlations, which are directly connected to ToF imaging,
cannot detect interactions in a localized system due to their
correspondence to particle transport. There is yet other information the ToF
reveals: One can also lower bound
the spatial entanglement of bosons in optical lattices \cite{Spatial}, building upon the ideas of constructing
quantitative entanglement witnesses \cite{quant-ph/0607167,Audenaert06,Guehne},
a notion of multi-partite entanglement $M(t)$ detecting a deviation from a best separable approximation,
as $M(t)\geq \max (0, \langle n\rangle  - \langle n(q)\rangle/|\hat{w}_0(q)|^2)$ for all $q$.
This quantity detects a reasonable notion of multi-particle entanglement, which is yet different from the bi-partite entanglement
discussed above. Since this measure is only onsite local,
we would expect that it cannot distinguish the long-range correlations of
an interacting disordered model from the dynamics inside the constant of
motion. This further motivates the quest to engineer appropriate
entanglement witnesses both accessible in optical lattice architectures
as well as probing key features of MBL, a quest that is in turn expected to
contribute to our understanding of MBL as such.

\subsection*{Acknowledgements}

We would like to thank the DFG (CRC 183, EI 519/14-1, EI 519/7-1, EI 519/15-1), the
European Commission (AQUS, SIQS, RAQUEL), the Templeton Foundation
and the ERC (TAQ)  for support.
This work has also received funding from the European Union's Horizon 2020
		research and innovation programme under grant agreement No 817482 (PASQuanS).
		We warmly thank A.\ Rubio Abadal, U.\ Schneider, C.\ Gross, I.\ Bloch,
A.\ Scardicchio, and R.\ Vasseur
for discussions. \emph{Note added:} This work was first submitted as a preprint as a blueprint for a joint
experimental-theoretical effort in progress. We now decided to properly publish this work in SciPost
as a scientific venue that is sympathetic to preprints. We insist that this work is still timely
and guiding present and future experiments.
  
\bibliographystyle{SciPost_bibstyle}

\appendix

\section{Numerical details}
\label{app:numerics}

In this appendix, we present the details of our numerical simulations.
Our results mainly rely on a matrix-product state simulation based on a TEBD code \cite{opentebd}, so an instance of a tensor network state
simulation. In order to corroborate the results, we have furthermore employed 
an exact diagonalization code \cite{scipy} that uses the particle number 
symmetry and keeps track
of the time evolution with a Runge-Kutta integration scheme.
For the non-interacting case, further checks were performed by 
an explicit simulation of the dynamical evolution of the covariance matrix, 
which takes
a particularly easy form in this case.

\begin{figure}[H]
  \centering
  \includegraphics[width=.6\textwidth]{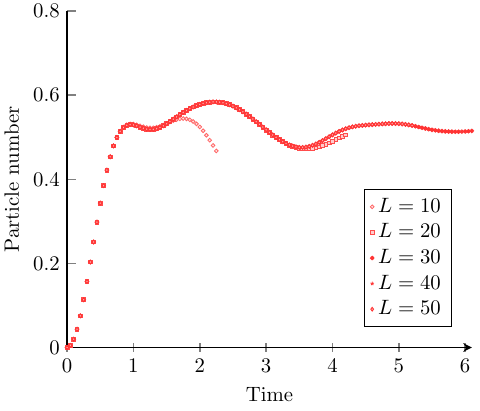}
  \caption{Finite size scaling for the evolution of particle density in the middle of
    the chain for a typical disorder realization.
    For $L=10,20$ an exact diagonalization code was used. The other system sizes are simulated
    with a TEBD code \cite{opentebd}.
  }\label{FS}
\end{figure}

For short times and the system sizes that can be achieved with exact diagonalization,
the codes agree up to a negligible error, thus also demonstrating that the chosen step size
in the $5$-th order Trotter decomposition used in TEBD \cite{opentebd} of $\tau_{\text{step}}$
does not produce significant errors.
This leaves only two potential error sources, the fact that numerics necessarily simulate
a finite system and the possibility of discarded weights accumulating over time.

\begin{figure}[H]
  \centering
  \includegraphics[width=.6\textwidth]{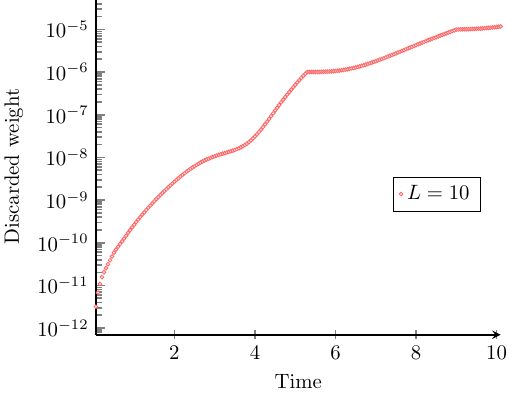}
  \caption{
    Evolution of the discarded weight. This plot varies strongly depending on the chosen
    disorder realization. From the $100$ realizations used for the averaged plots, the 
    realization with the largest discarded weights is shown here. 
  }\label{DW}
\end{figure}

Performing a finite size scaling, we find that comparably small systems are already
indistinguishable from the thermodynamic limit for the quantities considered here, see
Fig.\ \ref{FS}.
This is in agreement with the very slow growth of Lieb-Robinson cones expected in these
disordered systems.
To be on the safe side, we nevertheless carried out our numerical analysis on systems
with $L=80$ sites and open boundary conditions.

Having demonstrated that the considered system size is indistinguishable from the thermodynamic
limit only leaves the discarded weight as potential error source, see Fig.\ \ref{DW}.
The time evolution of this quantity, which is directly connected to spatial entanglement entropies,
depends strongly on the chosen disorder realization. In order to keep this discarded weight
small enough, we increase the bond dimension in the simulation in a three-step procedure up to
$d_{\mathrm{Bond}}=350$, which is sufficient to guarantee a discarded 
weight smaller than $2\cdot 10^{-5}$
for all disorder realizations.

\section{Bosonic model with on-site interactions}
\label{app:bosons}

In this appendix, we show additional simulation data for a measure similar 
to \textbf{Measure 2} for 
a related model that has been used in some of the experimental realizations
of MBL. This is the disordered Bose-Hubbard model with on-site interactions given by
\begin{equation}
  \label{eq:dis_bos_ham}
  H = \sum_j \left( b_j^\dagger b_{j+1} + \text{h.c.} \right) + \sum_j w_j n_j + U \sum_j n_j n_{j}  ,
\end{equation}
where $b_j$ denotes a bosonic operator on site $j$ and
$n_j=b_j^\dagger b_j$ is the local particle number operator and again
we draw the $w_j$ from the uniform
distribution on the interval $[-I,I]$.
In contrast to
the fermionic variant in the main text, we here need to restrict the local
Hilbert space in order to be able to perform numerics. We restrict
the local particle number to $k=3$ particles per site, but also made sure that 
enlarging the local dimension does not change our results qualitatively.
Moreover, our initial state is again an MIS state as defined in
Eq.\,\eqref{initial_state}, featuring an average particle number of 
0.5 per site.
The measure, we employ for bosons is identical to \textbf{Measure 2} with
the exception that the number operators were replaced by parity operators.
  \begin{measure}[Logarithmic information propagation] In order to examine the
  spatial spreading of parity-parity correlations, we define
  the quantity $P_{\mathrm{Corr}}(t)$,
  \begin{eqnarray}
    f_{\mathrm{Corr}}(k,t) &:=&   | \langle p_{L/2} p_{L/2+k} \rangle - \langle p_{L/2} \rangle \langle p_{L/2+k} \rangle |,\\
    P_{\mathrm{Corr}}(t) &:= & \sum_k f_{\text{Corr}}(k,t) k^2\,,
    \end{eqnarray}
    where $p$ is the local parity operator.
  \end{measure}

\begin{figure}
  \centering
  \includegraphics[width=0.6\textwidth]{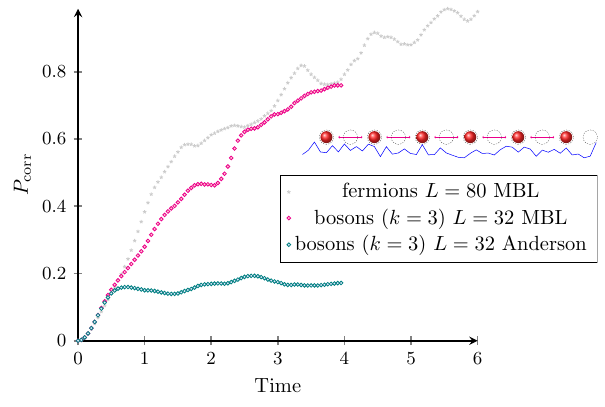}
  \caption[Numerical findings]{
    Plotted are the results of a TEBD simulation 
    of the dynamical evolution of the parity-parity correlations $P_{\mathrm{corr}}$.
    The initial state $\psi$ is again found in Eq.~\eqref{initial_state}
    under the Hamiltonian in Eq.~\eqref{eq:dis_bos_ham} for the case of an Anderson insulators with $U=0$ and
    MBL with $U=2$. We compared the results of the fermionic MBL setting and the bosonic MBL and 
    Anderson setting with a local Hilbert space dimension truncation $k=3$. 
    Every data point corresponds to an average over $100$ realizations.
  }
  \label{f:boson}
\end{figure}

In Fig.\,\ref{f:boson}, we show \textbf{Measure 4} for the Anderson
($U=0$) and MBL ($U=2$) case. Similarly to the main text, we find that in the
non-interacting case the measure saturates after few tunnelling times. 
In contrast, for the interacting model we find that the measure grows
in comparable fashion to the fermionic counterpart (grey stars).
This suggests that the correlation measure can be employed in similar
models as well.

\end{document}